\documentclass[pdflatex,sn-mathphys-num]{sn-jnl}

\usepackage{graphicx}%
\usepackage{multirow}%
\usepackage{amsmath,amssymb,amsfonts}%
\usepackage{amsthm}%
\usepackage{mathrsfs}%
\usepackage[title]{appendix}%
\usepackage{xcolor}%
\usepackage{textcomp}%
\usepackage{manyfoot}%
\usepackage{booktabs}%
\usepackage{algorithm}%
\usepackage{algorithmicx}%
\usepackage{algpseudocode}%
\usepackage{float}
\usepackage{listings}%

\theoremstyle{thmstyleone}%

\theoremstyle{thmstyletwo}%

\theoremstyle{thmstylethree}%

\begin{document}

\title[Article Title]{A Framework For Decentralized Micro-credential Verification Towards Higher Qualifications}

\author*[1]{\fnm{Abrar} \sur{Mahbub}}\email{abrar.tanim@gmail.com}
\author*[1]{\fnm{Humira} \sur{Saria}}\email{sariahumirai@gmail.com}
\author*[1]{\fnm{Md. Foysal} \sur{Hossain}}\email{foysalh1999@gmail.com}
\author*[1]{\fnm{Nafees} \sur{Mansoor}}\email{nafees.mansoor@ulab.edu.bd}
\equalcont{These authors contributed equally to this work.}

\affil*[1]{\orgdiv{Computer Science and Engineering}, \orgname{University of Liberal Arts Bangladesh}, \orgaddress{\state{Dhaka}, \country{Bangladesh}}}

\abstract{Student retention is one of the rising problems seen in educational institutions. With the rising cost of education and issues in the education sector, such as curriculum relevance, student engagement, and rapidly changing technological advancements, ensuring the relevance of academic programs in a fast-evolving job market has created a significant concern for educational institutions. With the intent to adapt to such challenges, educational institutions are dealing with alternative solutions for education, in which micro-credentials are at the very center of this, which are short-term academic programs or standalone courses.  However, one of the challenges of micro-credentials is a lack of credit transfer among institutions. With the lack of standardization of assessments among educational institutions, it is difficult to transfer micro-credentials to larger qualifications. Regarding such challenges, micro-credentials with blockchain technology can bring significant benefits. Blockchain technology offers a decentralized and immutable platform for securely storing and verifying credentials. This paper presents a prototype model for micro-credential verification. With the policies decided by the educational institution, the learner provides a micro-credential certificate to the system. Upon validation of the certificate by the verifying body, the educational institution will review the assessment criteria and provide exemptions based on the provided criteria. The prototype uses the Hyper-ledger Fabric platform and utilizes off-chain technology, which acts as a middle-man storage platform. With the combination of off-chain and on-chain technologies, congestion on the blockchain is reduced, and transaction speed is improved. In summary, this research proposes a prototype for secure micro-credential verification and a more efficient course exemption process.}

\keywords{Blockchain, Micro-credentials, Education, Decentralization, Verification, Off-chain, Academic }

\maketitle

\section{Introduction}

Education is a fundamental human right, significantly impacted by the pandemic’s global and digital shift. Although this has affected many areas of life, it has primarily impacted education and learning \cite{b1}. In a time when knowledge and skills need to be constantly updated, a three- or four–year degree may not be sufficient for many jobs and works \cite{b2}. Today's education requires updated,flexible, and personalized learning to keep up with constantly evolving abilities and expertise. The limitations of the current education system include outdated textbook content, rigid course offerings, and inflexible time constraints. 

Education aims to build a more inclusive, effective, and efficient system that can change with the evolving needs of society \cite{b1}. Nowadays, many students have dropped out of higher education. One of the main reasons for this, is the lack of flexibility and personalized learning opportunities. The current one-size-fits-all approach fails to accommodate individual learning preferences and paces. Even for those who graduate, many struggle to succeed in the fast-paced modern economy while keeping up with new knowledge and skills. Unfortunately, the current education system has failed to keep up with the digital age and adopt a more modern, flexible, and personalized approach.

To address this issue, the proposed system uses credit-bearing micro-credentials. Micro-credentials are learning outcomes from shorter educational or training activities; they are generally divided into two types, which are credit-bearing and non credit-bearing. The formal qualification frameworks of educational institutions do not officially govern micro-credentials. However, they offer formal guidance since learners want micro-credentials to be clear and transferable to formal credentials. Micro-credentials are typically offered online but can also be offered in an on-site or hybrid format and used in stand-alone courses or as part of an academic program. Some institutions accept micro-credentials for official certification, and they can be stacked toward the institution's credentials. Micro-credentials are expected to be clear and transferable to formal credentials.

The popularity of micro-credentials is increasing globally due to their flexible approach. Education and training organizations collaborate with others to create new programs and re-brand existing ones. To provide micro-credentials \cite{b3}, credit-bearing micro-credentials are available in official guidelines and relevant to formal qualifications. They may be accepted into official certification or degree programs and stack-able toward higher certifications. Micro-credential adoption in higher education (HE) comes in three forms: complete embrace, partial embrace, and minimum embrace. Moreover, why do we use credit-bearing micro-credentials ? Because micro-credentials are more compact, focused, and flexible than standard qualifications, they can address every one of the concerns with an education that we have discussed above \cite{b4}. According to experts, educational institutions that offer innovative programs and study options beyond the standard are ideal for micro-credentials to flourish. Additionally, increasing the utilization of micro-credentials may positively impact dropout rate \cite{b5}.

Millions of students graduate yearly and apply for jobs, but verifying these certificates as well as store or management, can be difficult. Moreover, there is always a risk of loss or damage to the certificate. Micro-credentials, especially crest-bearing ones, have increased the need for certificate verification. Digital certificates are prone to forging, although modern systems use encryption techniques to prevent fraud. However certification fraud is still common, and design tools can create convincing fake certificates. Physical certificates are also impractical and not environmentally friendly, with frequent issues like loss and damage. Additionally, if authentication is required, it creates a very disadvantageous situation 
 \cite{b6}.

The proposed system uses Blockchain, a relatively new technology, to overcome these issues. So, what are the advantages of blockchain technology? The three primary characteristics of the blockchain are immutability, traceability, and decentralization. This technology's decentralized features allow it to address issues of trust between strangers. Information manipulation is impossible due to its immutability. Because blockchain operates independently of a central authority (an intermediary), it offers the highest security and confidence. IPFS stores multimedia files, such as documents or images; the IPFS protocol creates a hash key and stores it in the Blockchain, offering greater security and efficiency\cite{b6}.

The proposed solution aims to design and create a decentralized, transparent certificate verification system using blockchain and a mix of off-chain and on-chain storage. The system will be used for academic certificate verification and credit-bearing micro credential verification. The system enables encryption and timestamp technology, which enable secure and impenetrable storage of certificates, reducing the worry of certificate damage or loss. Also, the blockchain has a cryptographic data structure that makes records virtually impenetrable, ensuring that data stored on the blockchain has not been tampered with or changed. Also, the proposed solutions have a user-friendly interface that helps students have seamless access.

This paper aims to determine the current knowledge and understanding of credit-bearing micro-credentials and blockchain technology. After the introduction, section 2 includes related papers and the literature review. Section 3 presents an architecture for how blockchain technology is used for academic and credit-bearing micro-credential certificate verification. Some subsections in section 3 include a workflow diagram, network diagram, etc. Section 4 contains Research Impact. Finally, the 5th section contains conclusion.

\section{Related Work}

The primary objective of this section is to understand the current state of knowledge regarding credit-bearing micro-credentials and certificate verification using blockchain technology. The objective is to understand the challenges associated in such related papers and understand the design approach towards their respective systems.

This study focuses on utilizing blockchain and IPFS technology to address the challenge of fake certification, which is very pressing issue in South Asia. With blockchain and IPFS the system aims to securely store and verify academic credentials.
The proposed system recognizes multiple stakeholder for this operation which are applicant, companies, admins and general user. Here applicants will upload their credentials which will be then verified by admins through contacting the HEI. After the credentials are verified they will be uploaded to IPFS where their hash will be encypted and securely stored in the blockchain. After they are secured companies can easily search for the credentials and verify their authenticity. 
With frameworks like ganache metamask, truffle, react and node js, the blockchain platform is created. the proposed solution offered in this study offers a transparent, efficient and secure method for verifying academic credentials \cite{b7}.\\

The research  proposes a blockchain based solution which deals with the inefficiencies that come with maintaining and certifying academic credentials through traditional method. The research highlights the potential of blockchain to provide a decentralized, secure and efficient solution. The author identifies 4 different stakeholders for the system which are accreditation body, university, students and employers. The system research utilizes Ethereum blockchain to implement smart contracts and uses IPFS to store the certificates. Moreover it uses the AES encryption method to store the certificates in the public blockchain. There are 3 main users to this system. First comes the issuer who is responsible for issuing the certificates and it can be universities or training centers. Secondly, the user comes which can be students or employers. The user will be able to share their information through a public key. And finally comes the accreditation body which validated the certificates \cite{b8}.\\

The research proposes a method to verify educational credentials and facilitate credit transfer which have been obtained locally or from any online educational portal. The research aims to enhance data accuracy and reliability of such credentials.The research proposes a token based framework to transfer credits using blockchain The credit value is transferred as token to the student's wallet after the completion of each course. The research presents a proof of work consensus mechanism to verify the tokens and is using the Ethereum blockchain to deploy smart contracts. The stakeholders in this system are the students and the universities. Through public key encryption and hashing techniques the security of the system is enhanced and the  scalability of the system is tested  through load testing  with multiple universities \cite{b9}.\\

The study proposes a system that is compatible with openbadges of IMS Global Learning Consortium.  OpenBadges supplies the badge setup, whereas IMS Global handles the badge certification. The article can be utilized for certificate issuance and management. It is also applicable for online or offline education and training courses. The education programs are broken up to micro learning units through which students can recieve digital badges upon completion. The process of the system is transparent when it is compared to other educational systems. Thus, in this study a blockchain based approached was taken and a block badge-based digital badge issuance and backpack management system was built. The IMS Global is used to set the standard for digital credentials through OpenBadge Specification platform. Mozilla OpenBadges framework is used to issue badges to students. Open badge can be set as a technical standard which is developed by IMS Global. Every digital badge will contain metadata. OpenBadge defines the format of badge meta data. Finally badgr is used for creating issuing and managing digital badges, whcih originally is not supported by blockchain. The article proposes to use blockchain in this process by utilizing the platform called Blockcert. The Blockcert platform is used to manage digital badges. When a digital badge is offered the issuing authority will record the transactions on the following network. After the transaction being recorded the badge repository which is called backpack in this system will serve as a centralized location for managing and accessing these digital badges. Two blockchain networks are used on this system which are the Bitcoin blockchain and the Ethereum network. For Bitcoin, the Ropsten test network is utilized, while for Ethereum, the Bitcoin server in regression test mode is employed. The REST API server is used regarding the operations of OpenBadge system and blockcert server is responsible for the transaction that are made in the blockchain \cite{b10}.\\

The paper introduces a global decentralized platform ecosystem built on blockchain technology.
Blockchain ensures secure, trusted, and decentralized
transactions without a central authority. EduCTX facilitates
managing, assigning, and presenting credentials for
individuals and educational institutions, enabling automation
of skill evaluation. It uses Ethereum blockchain, offering a
simplified, efficient environment to overcome language and
administrative barriers. The platform enhances access to
knowledge, addresses demographic challenges, reduces brain
drain, and improves employability and economy
competitiveness \cite{b11}.\\

The research proposes using blockchain technology in academic certification within higher education institutions and highlights the benefits and challenges of this process. The paper systematically reviews blockchain-based applications in education, focusing on academic certification, sovereign identity, and passport learning. It also discusses the differences between physical, digital, and blockchain certifications, steps to develop a decentralized application, and a use case diagram and framework for academic certification. Although they give a comprehensive framework for academic certification, they did not delve into micro-credentials\cite{b6}.\\

The paper proposes a reliable and tamper-proof certificate verification system using blockchain technology. It focuses on decentralizing the entire process of academic certificate insurance and verification to eliminate undesired behavior.
The proposed system harnesses the power of blockchain technology, specifically blockchain-based smart contracts, to automate the verification process. This not only enhances security and robustness but also eliminates single points of failure. All steps, from user registration to insurance and verification of certificates, are seamlessly executed on a blockchain using smart contracts on the Hyperledger Fabric Platform \cite{b12}.\\

This paper explores the potential of using blockchain and artificial intelligence (AI) to empower micro-credentials in the education sector. It highlights the importance of recognition and credit transfer for students studying at different institutions and settings. It emphasizes micro-credentials' role in providing easily accessible and transparent evidence of skills or knowledge. It also explains how blockchain technology can securely store and spread widely in micro-credentials while AI can facilitate their maintenance and distribution.
The paper suggests that educational blockchain implementations require permission from the blockchain owner, allowing employers to view data with the student's permission. It mentions several implementations of blockchain in education, such as Holberton School and Blockstack, which enable students to own their data. The paper also highlights the role of AI in supporting micro-credentials earned from decentralized and short-term courses and validating skills on the blockchain. It emphasizes the need for research into these technologies' technical and administrative integration, including developing standards for interoperability and acceptability among institutions and employers \cite{b13}.\\

This paper aims to design, implement, evaluate, and promote the adoption of a blockchain-based micro-credential system within a business school's executive education unit. The study aims to understand the value added by blockchain technology, the issuers and recipients of micro-credentials, and the challenges that must be addressed for widespread adoption. 
The paper employs a rigorous design science research methodology (DSRM) to conduct the study. DSRM encompasses six meticulous steps: problem identification, objectives definition, design and development, demonstration, evaluation, and communication. The researchers meticulously designed and implemented an independently managed blockchain-based micro-credential system within the university. The implementation is rigorously evaluated from the certificate issuer's and recipient's perspectives. However, their approach is only from a design or UI perspective \cite{b14}.\\

\section{Proposed System}
This section outlines the proposed system for certificate validation and course exemption. The system is based on blockchain technology. The general purpose for the system proposed is for certificate verification however, the focus is mainly on micro-credentials. Most specifically credit bearing micro-credentials. With the verification of credit-bearing micro-credentials the institutions will be able to provide course exemptions based on their policies.\\
Now regarding the exemption of courses, the respective educational institution will be driving factor to set up their policies. An educational institution based on their learning outcome will choose the micro-credentials required to get exemption from a course. Now this micro-credential courses can be standalone or stacked. Institutions joining the blockchain network can offer the micro-credential courses or they could collaborate with third party entities. \\
For the proposed system the platform Hyperledger Fabric is used. This platform is used as it allows fine-grained access control over who can access the data. For an application that needs to protect sensitive information, it is an essential feature. Moreover, this platform allows high scalability and performance which allows large number transactions per second. Another important feature of this platform is the deployment of smart contracts which is called chain-code in this platform. It can be written in general purpose language which makes it easier to create and deploy smart contracts.

\subsection{Proposed Framework}
\begin{figure}[h]
\caption{Workflow of the system}
\centering
\includegraphics[width=0.9\textwidth]{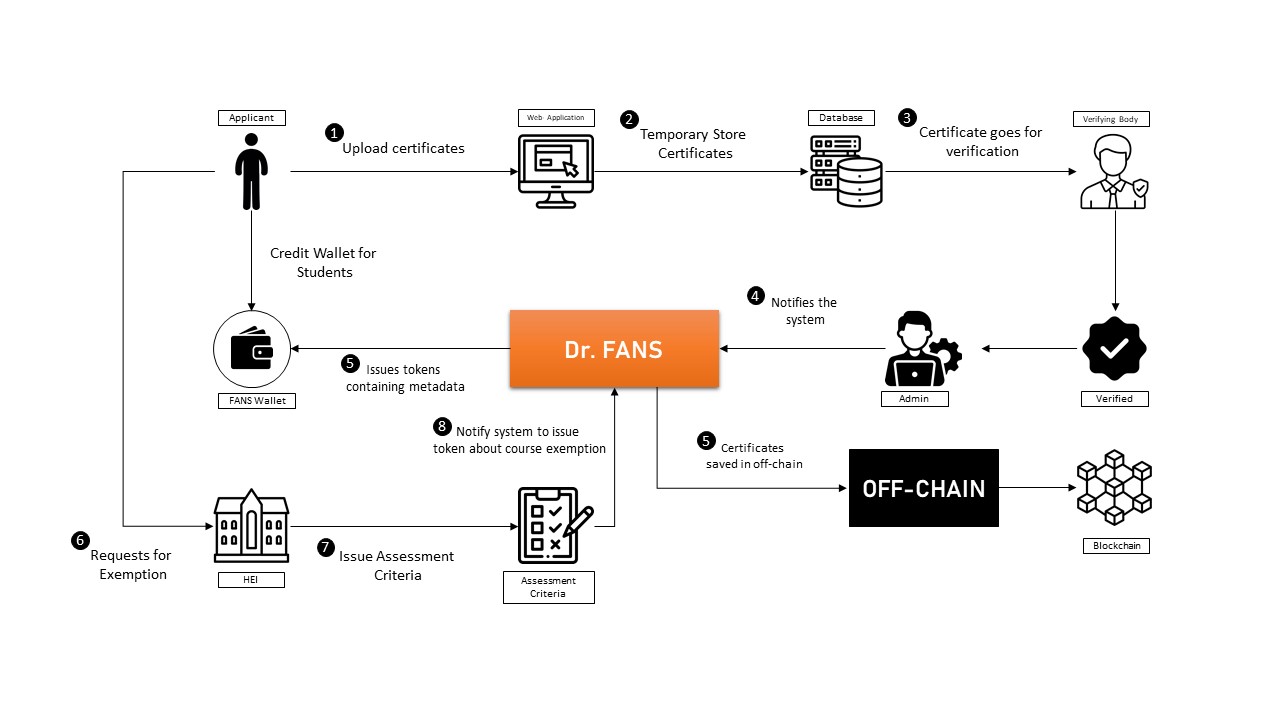}
\end{figure}
The workflow of the system is explained in figure 1. For the system the user will first have to register under their university name. After the account is verified and their personal wallet is created which is called FANS wallet, the user will be able to use the system. When a user wants to certify a course in the system. The user will login into the system and then will upload the credentials and necessary documents required through the web application. After the credentials are submitted, the documents are stored in a local database awaiting verification. Once the verification is completed by the verifying body, the admin of the system which is assigned by the institution will notify the system and will upload the documents for off-chain storage. For the off-chain storage, the documents are hashed and then stored in the blockchain. Alongside, a token will be transferred to the FANS wallet of the user containing metadata regarding the course. Upon receiving the token, the user will request an exemption towards the educational institutions using their tokens as evidence. Upon reviewing the tokens, the institution will issue an assessment criteria in the system and upon fulfilling the assessment criteria, the institution will give exemption.

\subsection{Network Overview}
Blockchain network serves as the foundation of the system. Blockchain networks are decentralized ledger that records transaction across a network of computer. With each transaction, the transaction is grouped into a block and are then linked in chronological order in order to form a chain. \\
\begin{figure}[h]
\caption{Network Overview}
\centering
\includegraphics[width=0.8\textwidth]{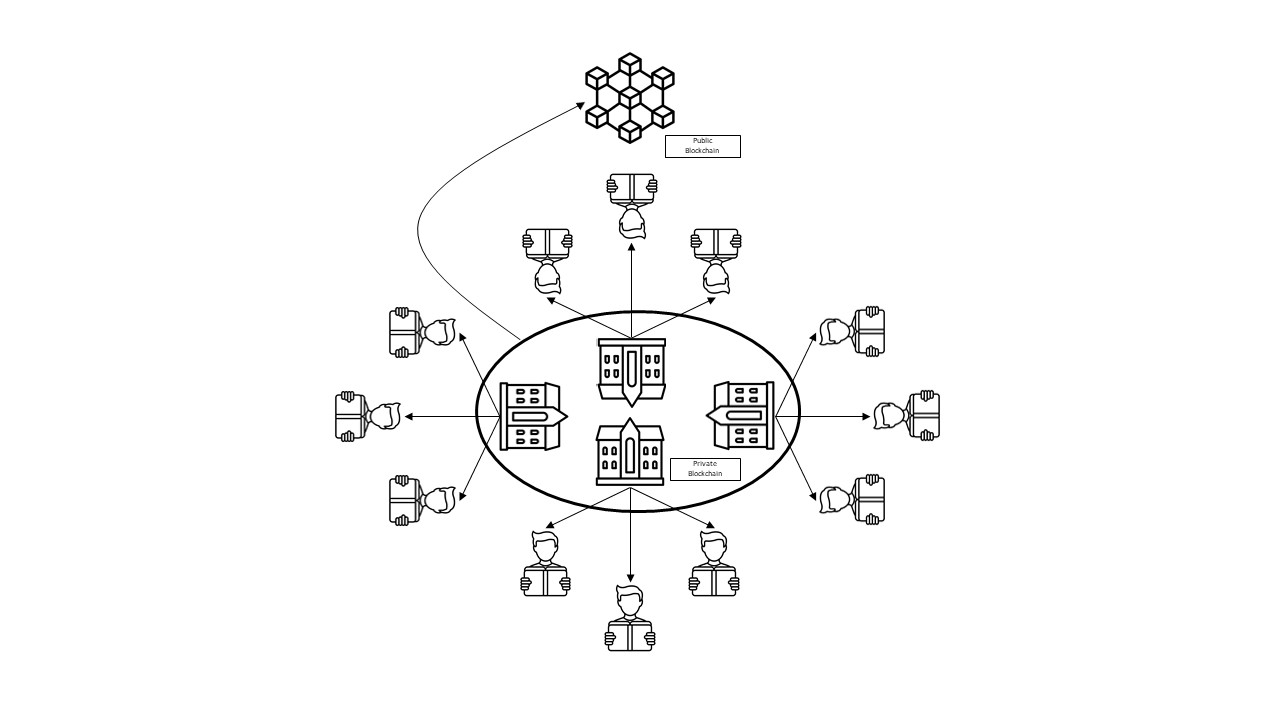}
\end{figure}

Now regarding the blockchain, three types of network are mostly present. They are public, permissioned and consortium blockchain. In a public blockchain, anyone can access the data while in a permissioned blockchain, only users with permission can access it. In a permissioned chain only a single entity acts as an authoritarian node. However, in a consortium blockchain, which is a mix of public and permissioned networks, there are multiple entities acting as authoritarian node. With multiple authorative nodes, it is assumed that there are trust to some degree, as any faulty transaction can be tracked down easily. Therefore, a lower computation power consensus algorithm can be used in such a case, which will increase the transaction throughput essentially in some degree. \\
In our system, we have proposed a consortium blockchain. Figure: 2 shows the network overview of our system. For a student to be a part of the network, the respective educational institution has to be a part of the permissioned network. By being part of the permissioned network, the institution will have the right to participate in the consensus. However, public blockchain is used in the case of certificate storage and hash storage. Upon the permission of the authorative nodes, the public blockchain is updated. The blockchain includes the certificates and also the FANS wallet. It is done such that users can access it at any time, and for various reasons, third party entities can also access it for verification of credentials. Even though these blockchains are public, they are securely encrypted and can be accessed only by the users persmission.

\subsection{Chain-Code}
Blockchain is a distributed ledger technology that allows transactions to be recorded over a network of computers in a safe, transparent, and immutable manner. Within the blockchain platform a contract can be initiated which is self-executing with the terms of agreement written in the code. It allows for the automation of complex transactions and processes without the need of intermediaries. One of the platforms for blockchain is called Hyperledger Fabric. Hyperledger Fabric supports the use case of consortium blockchain and creates a scalable environment for developers. Hyperledger fabric also has a self executing contract which is called chain-code. It is responsible for implementing transaction logic, data validation and state management for smart contracts. 
\begin{table}[h]
    
    \begin{tabular}{|l|} 
 \hline
\textbf{When a user applies for course certification} \\
 \hline
 Check if applicantID Exists Then: \\
    \quad transaction = createTransaction(issuerIdentity,applicantID, metadata)\\
    \quad signedTransaction = signTransaction(transaction, issuerPrivateKey, applicantPrivateKey)\\
    \quad submitTransaction(signedTransaction)\\
    \quad Brodcast in PerBC:\\
        \quad \quad consensusResult(signedTransaction)\\
    \quad IF consesusResult is TRUE:\\
        \quad \quad addToken(signedTransaction)\\
        \quad \quad validCert = encrypt(hash(signedTransaction))\\
        \quad \quad certificationBlock(applicantID, issuerID, validCert)\\
        \quad \quad hashCertificationBlock = hash(certificationBlock)\\
        \quad \quad Add certificationBlock to off-chain \\
        \quad \quad Add hashCertificationBlock to Public blockchain\\
        
    ELSE:\\
        \quad Send ("Application Rejected !!!")\\

ELSE:\\
    Send ("Account Not Found !!!")\\
    \hline
\end{tabular}
    \caption{Smart Contract Pseudocode for course certification}
    \label{tab:my_label}
\end{table}

Table: 1 showcases the pseudocode for the chain-code that is used during the certificate validation process.
When a user applies for certificate validation. The user ID is check with platform. If it exists then the chain-code is executed else the user is not notified with the issue. Upon successful account verification a transaction is created which is signed by the issuer which is the institution that the user belongs to and the user's private ID. After the transaction is signed, it is brodcasted in the public blockchain where the authoritarian nodes participate in the consensus of its legitimacy. Upon a successful consensus, the signed transaction is transferred to the student's wallet and the certificate is encrypted, hashed and safely moved to the off-chain for storage. The hash is also moved to blockchain to ensure immutability of the process.

\section{Research Impact}
In this modern age information is easily accessible then ever before. In the face of this higher educational institutions are facing various internal and external pressure. One of the problems that educational institutions are facing are student retention. It can no longer be assumed that students are achieving their learning objectives which are specified by the courses or programs. Student's evaluation regarding their progress towards learning outcomes should be done in a quantifiable way in order to be accredited. Moreover with the rising cost of education, learners are eager to secure employment quickly \cite{b15}. Another important factor is the recognition of prior learning (RPL). RPL allows individuals to receive credit of certification, knowledge and skills they have already acquired \cite{b16}. Such recognition reduces duplication of efforts in larger qualifications and motivates learners to engage in life-long learning. The system proposed in this research aims to combat such challenges in the education field. Moreover with such system, learners can have their own personalized learning paths and focus on areas that are most relevant to their goals and interests while moving towards their institutional qualifications.

\section{Conclusion}

This work proposes a possible solution for education that could solve the current education landscape's many concerns. Although there is also verification of credentials for academic and credit-bearing micro-credentials we discussed. In addition, prototypes will be made to test the proposed architecture. We also discussed credit-bearing micro-credentials, which have the potential to revolutionize education. It can solve many concerns of the current education system and provides a flexible approach.

Blockchain technology has immense potential for use in many sectors, such as finance, forensics, healthcare, agriculture, government, education, and so on. It can provide solutions to multiple sectors' problems.

Blockchain's usage in the education sector is immense, and in this globalized world, the education sector needs blockchain. Blockchain can be used in different academic areas. One of the most significant advantages of blockchain is that it is reliable, secure, impenetrable, and safe to store, and it does not use any intermediaries for transactions \cite{b6}. It uses decentralized architecture, which makes it trustworthy and unbiased. It has traceability, which ensures transparency. Its immutability features ensure that there is no manipulation. For future work, we plan to build a network to collaborate with industry and higher education institutes to provide more accessible, up-to-date, skill-based, and flexible education, thereby revolutionizing the education sector.

%
%

%
%
%
%

\end{document}